\documentclass[apj]{emulateapj}
\usepackage{mathptmx}

\def\gtorder{\mathrel{\raise.3ex\hbox{$>$}\mkern-14mu
             \lower0.6ex\hbox{$\sim$}}}
\def\ltorder{\mathrel{\raise.3ex\hbox{$<$}\mkern-14mu
             \lower0.6ex\hbox{$\sim$}}}

\slugcomment{Draft of \today}

\shorttitle{Space and velocity distributions of Neutron stars}

\shortauthors{Ofek}

\begin{document}

\title{Space and velocity distributions of Galactic isolated old Neutron stars}
\author{Eran~O.~Ofek\altaffilmark{1}}

\altaffiltext{1}{Division of Physics, Mathematics and Astronomy, California Institute of Technology, Pasadena, CA 91125, USA}

\begin{abstract}

I present the results of Monte-Carlo orbital simulations of
Galactic Neutron Stars (NSs).
The simulations take into account the up-to-date observed NS space and
velocity distributions
at birth, and account for their formation rate.
I simulate two populations of NSs. Objects in the first population were born
in the Galactic disk at a constant rate,
in the past 12\,Gyr. Those in the second population were formed
simultaneously 12\,Gyr ago in the Galactic bulge.
I assume that the NSs born in the Galactic disk
comprise 40\% of the total NS population.
Since the initial velocity distribution of NSs is not well known,
I run two sets of simulations, each containing $3\times10^{6}$
simulated NSs.
One set utilizes a bimodal initial velocity distribution
and the other a unimodal initial velocity distribution,
both are advocated based on pulsars observations.
In light of recent observational results,
I discuss the effect of dynamical heating by Galactic structure
on NS space and velocity distributions and show it can be neglected.
I present catalogue of simulated NS space and velocity vectors
in the current epoch, and
catalogue of  positions, distances and proper motions
of simulated NSs, relative to the Sun.
Assuming there are $10^{9}$ NSs in the Galaxy,
I find that in the solar neighborhood
the density of NSs is about $2-4\times10^{-4}$\,pc$^{-3}$,
and their scale height is about $0.3-0.6$\,kpc
(depending on the adopted initial velocity distribution).
These catalogue can be used
to test the hypothesis that
some radio transients are related to these objects.

\end{abstract}


\section{Introduction}
\label{sec:Introduction}

The present day space and velocity distributions of
Galactic isolated old Neutron Stars (NS),
are the subject of many studies.
Among the reasons for deriving the positional and kinematical properties
of isolated old NSs
was the suggestion that Gamma-Ray Bursts (GRBs)
may originate from Galactic NSs (e.g., Mazets et al. 1980).
This was
refuted\footnote{At least in the sense that the
majority of GRBs are extragalactic; see Kasliwal et al. (2008).}
by the homogeneous sky distribution
of GRBs (e.g., Meegan et al. 1992), and later on
by the discovery of their cosmic origin
(e.g., Metzger et al. 1997; van Paradijs et al. 1997).

Another exciting possibility,
that re-ignited these efforts,
was to detect isolated old NSs
in soft X-ray radiation
that may be emitted as they slowly accrete matter
from the interstellar medium (ISM; Ostriker, Rees \& Silk 1970;
Shvartsman 1971).
Treves \& Colpi (1991), and Blaes \& Madau (1993)
predicted that $\sim10^{3}$--$10^{4}$ isolated old
NS accreting from the ISM would be detected by the
{\it ROSAT} Position Sensitive Proportional Counters
(PSPC) all sky survey (Voges et al. 1999).
Although intensive searches for these objects
were carried out
(e.g., Motch et al. 1997; Maoz, Ofek, \& Shemi 1997; 
Haberl, Motch, \& Pietsch 1998;
Rutledge et al. 2003;  Ag{\"u}eros et al. 2006),
only a handful of candidates were found.
However, these are presumably young cooling NSs,
rather than old NSs whose luminosity
is dominated by accretion from
the ISM (e.g., Neuh\"{a}user \& Tr\"{u}mper 1999; Popov et al. 2000; Treves et al. 2001).
Apparently, the reasons for the rareness of these
objects in the {\it ROSAT} source catalog is
that their typical velocities were underestimated
by earlier studies (i.e., Narayan \& Ostriker 1990). 
It is also possible that their accretion rate is
below the Bondi-Hoyle rate\footnote{Bondi \& Hoyle (1944).}
(e.g., Colpi et al. 1998; Perna et al. 2003; 
Toropina et al. 2001, 2003, 2005; see however Arons \& Lea 1976,1980),
or that these NSs are in an ejection stage
(e.g., Colpi et al. 1998; Livio et al. 1998; Popov \& Prokhorov 2000).

During the last several years, new classes of radio
transients were found
(e.g., Bower et al. 2007; Matsumura et al. 2007; Niinuma et al. 2007; Kida et al. 2008;
see also Levinson et al. 2002; Gal-Yam et al. 2006).
Bower et al. (2007) found several examples of these transients
in a single small field of view,
and showed that they have time scales above 20\,min but below seven days,
and they lack
any quiescent X-ray, visible-light, near-Infrared (IR), and radio
counterparts.
Recently, Ofek et al. (2009) suggested that these transient
events may be associated with Galactic isolated old NSs.
However, testing this hypothesis requires
knowledge of the NSs space distribution.

There are two approaches in the literature for calculating
the theoretical space and velocity distributions of old NSs.
Paczynski (1990), Hartmann, Woosley, \& Epstein (1990),
Blaes \& Rajagopal (1991), Blaes \& Madau (1993),
and Posselt et al. (2008) 
carried out Monte-Carlo simulations of NS orbits.
In such simulations, the positions
and velocities of NS at birth
are integrated assuming some
non-evolving Galactic potential.
A second approach is to use some sort of semi-analytic approximation
in order to estimate the ``final'' NS space and velocity distributions.
Frei, Huang, \& Paczynski (1992) calculated the
final vertical density and velocity distributions of NSs,
assuming that the gravitational potential
is a function only of the height, $z$, above the Galactic plane.
Using the epicyclic approximation,
Blaes \& Rajagopal (1991) developed a technique
that allows calculation of the full three dimensional velocity
distribution of NSs.
However, they showed that this method is not adequate
for fast moving objects, which constitute the majority
of the NS population.
Blaes \& Madau (1993)
used the thin-disk approximation to calculate the space and velocity
distributions of NSs.
In this prescription, the radial motion of NSs
are controlled by the Galactic
potential in the Galactic disk, regardless of
the vertical height, $z$, of the NS above/below the Galactic plane.
Finally,
another solution was presented by Prokhorov \& Postnov (1994),
who assume that the ergodic hypothesis is correct
(see discussion in Binney \& Tremaine 1987).

Madau \& Blaes (1994) noted that all these approaches neglect dynamical
heating of NSs due to encounters with giant molecular clouds, spiral arms, and
stellar ``collisions'' (e.g., Kamahori \& Fujimoto 1986, 1987;
Barbanis \& Woltjer 1967; Carlberg \& Sellwood 1985; Jenkins \& Binney 1990).
They crudely estimated the order of magnitude of this effect
by applying the force-free diffusion
equation to the vertical height, $z$, and NS speed distributions.
They found that, at the solar neighborhood,
dynamical heating may decrease
the local density of NSs by a factor of $\sim1.5$
relative to a non-heated population.

In this paper I present the results of Monte-Carlo orbital simulations
of Galactic NSs.
These simulations improve upon past efforts (e.g., 
Paczynski 1990; Hartmann, Epstein, \& Woosley 1990,
Blaes \& Rajagopal 1991; Blaes \& Madau 1993;
Popov et al. 2005; and Posselt et al. 2008)
in several aspects.
First, I use up-to-date space and velocity distributions
of NSs at birth from Arzoumanian et al. (2002)
and Faucher-Gigu\`{e}re \& Kaspi (2006).
Second, I assume a birth rate of NSs along the Galaxy life time,
instead of assuming that all the NSs were born about 10~Gyr ago.
Third, I generate a large sample of simulated NSs,
which is about one to two orders of magnitude larger
than those presented by previous efforts.

The structure of this paper is as follows:
In \S\ref{sec:ModelIng} I present
the model ingredients for the Monte-Carlo simulations.
In \S\ref{sec:DynHeat} I discuss dynamical heating
and show that it can be neglected.
The Monte-Carlo simulations are described in \S\ref{sec:MC},
and the catalogue of simulated NSs are presented in
\S\ref{sec:Cat}.
The results of these simulations are presented in \S\ref{sec:NSres},
and in \S\ref{sec:Disc} I summarize the results and compare them with
some of the previous efforts.

\section{Model ingredients}
\label{sec:ModelIng}

In the following subsections, I present
the ingredients of the Monte-Carlo simulations.
These are: NSs birth rate (\S\ref{sec:NSbirthrate});
space and velocity distributions of NSs at birth (\S\ref{sec:NSbirthdist});
and the Galactic potential (\S\ref{sec:GalPot}).

\subsection{NSs birth rate}
\label{sec:NSbirthrate}

The stellar population in the Milky Way is composed
of at least two major components:
a bulge and a disk.
I therefore,
simulate two populations of NSs. A Galactic disk-born population
and a Galactic bulge population.
For the Galactic-disk NSs, I assume
a continuous constant formation
rate in the past 12\,Gyr.
This assumption is motivated by the analysis
of Rocha-Pinto et al. (2000) that did~not find
any major trends in the star formation rate in
the disk of our Galaxy.
I assume that the Galactic-bulge NSs
were born in a single burst 12~Gyr ago
(Ferreras, Wyse, \& Silk 2003; Ballero et al. 2007; Minniti \& Zoccali 2008).

As I discuss in this section,
the simulations presented here consist of $40\%$ disk-born NSs and $60\%$
bulge-born NSs.
The actual ratio of disk to bulge NSs
is unknown and may be significantly different
than the one assumed here.
Therefore, in the resulting catalogue (\S\ref{sec:Cat})
I specify the origin of each simulated NS
and present the results also for each population separately.

The total number of NSs in the Milky Way is constrained by
the chemical composition of the Galaxy.
Specifically, the iron content of the Galaxy implies that the total number of
core collapse supernovae (SNe; and therefore NSs) that exploded
in the Milky Way is about $10^{9}$ (Arnett, Schramm, \& Truran 1989).
However,
the star formation rate in the Galactic disk
was approximately constant
during the last $\sim12$\,Gyr
(Rocha-Pinto et al. 2000; see also Noh \& Scalo 1990).
Therefore, with the current NS birth rate of up to one in 30~years
(e.g., Diehl et al. 2006),
one expects that there will be $\ltorder4\times10^{8}$
disk-born NSs in our Galaxy.
The predictions based on the SN rate ($\ltorder4\times10^{8}$)
and the chemical evolution  ($\sim10^{9}$) are therefore
inconsistent.
Ways around this problem include uncertainty in the
SN rate (e.g., Arnett et al. 1989), and difficulty in
estimating the star formation history of our Galaxy
(Keane \& Kramer 2008).
Another viable resolution of this discrepancy is
that the Galactic bulge had a higher star formation
rate at earlier times.

The Galactic bulge contains a considerable number of
stars that were born,
apparently, in a $\sim1$\,Gyr-long, burst
about 12\,Gyr ago
(Ferreras, Wyse, \& Silk 2003; Ballero et al. 2007; Minniti \& Zoccali 2008).
I therefore, assume here that in addition to the disk-born
NSs there are up to
$6\times10^{8}$ NSs that were formed in the Galactic bulge
about 12\,Gyr ago.
This number was selected such that the total number of NSs in
the Galaxy is $10^{9}$.

I note that
the Galactic bulge contains only about 20\% of the
Galactic stars (e.g., Klypin et al. 2002).
However, Ballero et al. (2007) argued that the initial mass
function, $dN/dM$, of the bulge is skewed toward high masses,
with $x$, the power-law index in the
stellar mass function ($dN/dM\propto M^{-(1+x)}$; e.g., Salpeter 1955),
being $<0.95$.
Assuming that stars with masses exceeding 8\,M$_{\odot}$
produce NSs within a relatively short time
after their birth,
this suggests that the number of NSs born in the Galactic bulge
is a few times larger than the number of NSs born in the Galactic
disk, per unit stellar mass.
Therefore, the assumption
that the majority of NSs were born in the Galactic
bulge is in rough agreement
with the expected fraction of high mass ($\gtorder8$\,M$_{\odot}$)
stars in the bulge, relative to the disk.
Nevertheless, as stated before, the ratio between the numbers of
bulge-born to disk-born NSs is uncertain.
The simulations presented here
cover a wide range of possibilities regarding birth place and time
scenarios.

\subsection{NSs space and velocity distributions at birth}
\label{sec:NSbirthdist}

Following, Paczynski (1990), I adopt a radial birth
probability distribution, $p(R)$,
which follows the exponential stellar disk:
\begin{equation}
dp(R)=a_{R}\frac{R}{R_{d}^{2}}\exp{\Big(-\frac{R}{R_{d}}\Big)}dR,
\label{ExpDisk}
\end{equation}
where $R$ is the distance from the Galactic center
projected on the Galactic plane,
$R_{d}=4.5$\,kpc is the exponential disk scale length, 
and $a_{R}$ is the normalization given by:
\begin{equation}
a_{R}=[1-e^{-R_{max}/R_{d}} (1 + R_{max}/R_{d})]^{-1},
\label{ExpDiskNorm}
\end{equation}
where $R_{max}=15$\,kpc, is the disk truncation radius.
Yusifov \& K\"{u}\c{c}\"{u}k (2004) estimated the radial
distribution (measured from the Galactic center) 
of Galactic pulsars, and found it to be more concentrated than
predicted by the Paczynski (1990) distribution.
However, I note that the supernova remnant
Galactic radial distribution
is in good agreement with the Paczynski (1990) model.

For the NS bulge population I assume that the space
distribution at birth has the same functional
form and parameters as the NS disk population,
but with $R_{d}=1$\,kpc in Eqs.~\ref{ExpDisk}
and \ref{ExpDiskNorm}.

For the initial velocity probability distribution,
I use two different models.
Both models are consistent with the observed
velocity distribution of young pulsars.
One is a bimodal velocity distribution composed of
two Gaussians found by Arzoumanian et al. (2002),
and the second is
based on a double-sided exponential unimodal velocity distribution
(Faucher-Gigu\`{e}re \& Kaspi 2006).

{\bf Bimodal velocity distribution:}
The first initial velocity distribution we used is from 
Arzoumanian et al. (2002).
It consists of two components.
The three-dimensional speed, $v$, probability density of this model is:
\begin{equation}
p(v) = 4\pi v^{2} \sum_{j=1,2}{ \frac{f_{j}}{(2\pi\sigma_{v,j}^{2})^{3/2}}\exp{(-v^{2}/[2\sigma_{v,j}^{2}])}}
\label{BirthVelDist2}
\end{equation}
where $j\in \{1,2\}$,
$f_{1}=0.40$ and $f_{2}=0.60$ are the fractions of the two
NS velocity components, $\sigma_{v,1}=90$\,km\,s$^{-1}$,
and $\sigma_{v,2}=500$\,km\,s$^{-1}$.

{\bf Unimodal velocity distribution:}
The second model I use for the velocity distribution
of NSs at birth was advocated
by Faucher-Gigu\`{e}re \& Kaspi (2006).
In their model, each one-dimensional (projected)
velocity component, $v_{i}$, consists of a double-sided exponential
probability distribution:
\begin{equation}
p(v_{i}) = \frac{1}{2\langle v_{i}\rangle} \exp{\Big( -\frac{|v_{i}|}{\langle v_{i}\rangle} \Big)},
\label{BirthVelDist1}
\end{equation}
where $\langle v_{i}\rangle=180$\,km\,s$^{-1}$ is
the mean velocity in each direction, $i$.
I note that drawing the velocity components of simulated NSs
directly from Eq.~\ref{BirthVelDist1} will result in an aspherical
distribution.
In order to avoid this problem the
corresponding three-dimensional distribution is required.
Faucher-Gigu\`{e}re \& Kaspi (2006) noted that this
distribution is difficult to derive
analytically, and did~not provide its functional form.
However, it is possible to show that
the corresponding spherically symmetric three-dimensional
distribution corresponding to Eq.~\ref{BirthVelDist1} is
given by
\begin{equation}
p(v) = \frac{v}{\langle v_{i} \rangle^{2}} \exp{\Big( -\frac{|v|}{\langle v_{i}\rangle} \Big)}.
\label{BirthVelDist1_3}
\end{equation}

\begin{figure}
\centerline{\includegraphics[width=8.5cm]{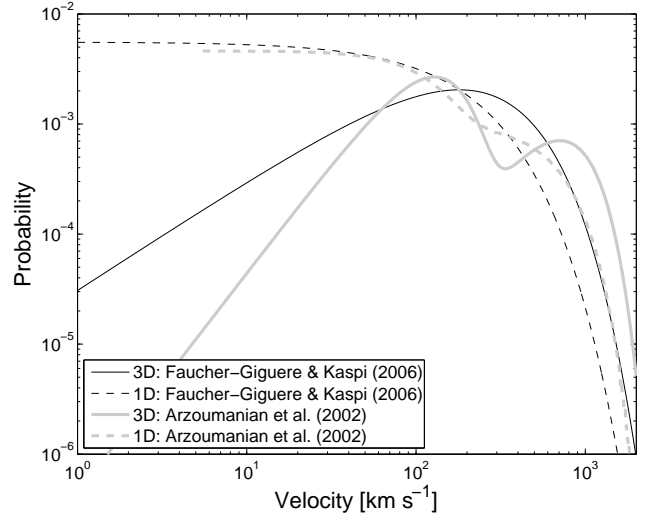}}
\caption{The initial velocity distribution of NSs at birth
relative to a rotating galaxy.
The {\it dashed} lines represent the one-dimensional (1D)
velocity distributions, while the {\it solid} lines
show the three-dimensional (3D) speed distributions.
The {\it black thin} lines are for the 
Faucher-Gigu\`{e}re \& Kaspi (2006) unimodal
double-sided exponential velocity distribution,
while the {\it gray thick} lines show
the Arzoumanian et al. (2002)
bimodal Gaussian velocity distribution.
\label{fig:InitVelDist_Compare}}
\end{figure}

In Figure~\ref{fig:InitVelDist_Compare}
I show the one-dimensional ({\it dashed} lines) and
three-dimensional ({\it solid} lines)
initial velocity probability distributions
based on
Faucher-Gigu\`{e}re \& Kaspi (2006) velocity distribution
({\it thin black} lines) and
the Arzoumanian et al. (2002)
distribution ({\it thick gray} lines).
%
%
Several other fits for the pulsars velocity distribution
at birth are available (e.g., Hansen \& Phinney 1997; Cordes \& Chernoff (1998); Hobbs et al. 2005; Zou et al. 2005).
However, the Arzoumanian et al. (2002)
and  Faucher-Gigu\`{e}re \& Kaspi (2006) velocity distributions
cover a wide range of possibilities, that reflect
the current uncertainties in the pulsars velocity distribution
at birth.

The two models are considerably different from each other,
mainly in the low and high velocity tails.
Relative to the
Arzoumanian et al. (2002) velocity distribution
the Faucher-Gigu\`{e}re \& Kaspi (2006) 
distribution has
excess of NSs at low velocities.
Since the slowest NSs are the most efficient accretors
from the ISM (e.g., Ostriker et al. 1970),
these differences may have a significant impact
on the detectability of isolated old NSs
in X-ray wavebands (e.g., Blaes \& Madau 1993).

The velocity distributions,
in Eqs.~\ref{BirthVelDist2} and \ref{BirthVelDist1_3},
are given relative to
a rotating disk.
Therefore, I added the disk rotation velocity to these birth velocities
\begin{equation}
v'_{x} = v_{x} - V_{circ}(R)\sin(\theta),
\label{VreldiskX}
\end{equation}
\begin{equation}
v'_{y} = v_{y} + V_{circ}(R)\cos(\theta),
\label{VreldiskX}
\end{equation}
\begin{equation}
v'_{z} = v_{z},
\label{VreldiskX}
\end{equation}
where $v_{i}$ is the velocity vector relative to the rotating disk
(i.e., the one obtained from
Eqs.~\ref{BirthVelDist2} or \ref{BirthVelDist1_3}),
$v'_{i}$ is the velocity vector in a non-rotating (inertial)
reference frame,
$R = (x^{2}+y^{2})^{1/2}$,
$\theta$ is azimuthal angle on the Galactic disk
($={\rm arctan2}[y,z]$), which was selected from
a uniform random distribution,
and $x$, $y$, and $z$ are the position in a Cartesian,
non-rotating, coordinate system,
whose origin
is the Galactic center, and $z$ is vertical to the Galactic plane.
Finally, $V_{circ}$ was obtained from
\begin{equation}
V_{circ}(R) = \sqrt{-R \nabla_{R}{\Phi}},
\label{Vcirc}
\end{equation}
where 
$\Phi$ is the Galactic potential (see \S\ref{sec:GalPot}),
at the point of interest.

Finally, I assume that the vertical height above the Galactic
mid-plane of NSs at birth, $z$, is drawn from a Gaussian distribution
($\propto \exp[-z^{2}/(2\sigma_{b})]$),
with $\sigma_{b}=0.16$\,kpc and $0.05$\,kpc,
for the Arzoumanian et al. (2002)
and Faucher-Gigu\`{e}re \& Kaspi (2006) initial velocity distributions,
respectively.

\subsection{The Galactic potential}
\label{sec:GalPot}

Following Paczynski (1990), I assume
that the Galactic potential is composed of
a disk component, an spheroidal component,
and an halo component.
I also added a central black hole component.

The disk and spheroid components are described by the
following potential proposed by Miyamoto \& Nagai (1975),
\begin{equation}
\Phi_{k}(x,y,z)=-\frac{GM_{k}}{ \Big\{x^{2}+y^{2}+[a_{k}+(z^{2}+b_{k}^{2})^{1/2} ]^{2} \Big\}^{1/2}},
\label{potential_disk_sphere}
\end{equation}
where $a_{k}$, $b_{k}$ and $M_{k}$ are given in Equations~\ref{eq:Potential_d_par}
and \ref{eq:Potential_s_par},
and where $k=d$ corresponds to the disk component and $k=s$ corresponds to
the spheroid component, and $G$ is the Gravitational constant.
Next, the halo potential is given by
\begin{equation}
\Phi_{h}(r)=-\frac{GM_{h}}{r_{h}} \Big[ \frac{1}{2} \ln{ \Big(1+\frac{r^{2}}{r_{h}^{2}} \Big)} + \frac{r_{h}}{r} \arctan{ \Big(\frac{r}{r_{h}} \Big) } \Big],
\label{potential_halo}
\end{equation}
were $r^{2}=x^{2}+y^{2}+z^{2}$, $M_{h}$, and $r_{h}$ are listed in Eq.~\ref{Potential_h_par}.
Furthermore, I added a component representing the Galactic
central massive black-hole (e.g., Ghez et al. 1998):
\begin{equation}
\Phi_{bh}(r)=-\frac{GM_{bh}}{r},
\label{potential_halo}
\end{equation}
where $M_{bh}$ is given in Eq.~\ref{Potential_bh_par} (Eisenhauer et al. 2005).
The choice of parameters listed here reproduces
the observed Galactic rotation, local density, and local column density (see Paczynski 1990 for details):
\begin{eqnarray}
a_{d}=3.7~{\rm kpc},~ &  b_{d}=0.20~{\rm kpc},~  & M_{d}=8.07\times10^{10}~{\rm M_{\odot}},
\label{eq:Potential_d_par}
\end{eqnarray}
\begin{eqnarray}
a_{s}=0.0~{\rm kpc},~ &  b_{s}=0.277~{\rm kpc},~  & M_{s}=1.12\times10^{10}~{\rm M_{\odot}},
\label{eq:Potential_s_par}
\end{eqnarray}
\begin{eqnarray}
r_{h}=6.0~{\rm kpc},~ & M_{h}=5.0\times10^{10}~{\rm M_{\odot}},
\label{Potential_h_par}
\end{eqnarray}
\begin{equation}
M_{bh}=3.6\times10^{6}~{\rm M_{\odot}}.
\label{Potential_bh_par}
\end{equation}
I note that the black hole contributes
about $4\%$ ($0.6\%$) of the gravitational potential at distance of 1\,pc (10\,pc) from the black hole.
Therefore, its influence on the Galactic potential is
negligible and it does~not
change the fitted parameters in Eqs.~\ref{eq:Potential_d_par}-\ref{Potential_bh_par}.
However, it may heat NSs passing nearby.
Finally, the Galactic potential is the sum of these four components:
\begin{equation}
\Phi = \Phi_{d} + \Phi_{s} + \Phi_{h} + \Phi_{bh}.
\label{TotalPotential}
\end{equation}

\section{Dynamical Heating}
\label{sec:DynHeat}

Dynamical heating, presumably by giant molecular clouds
(e.g., Kamahori \& Fujimoto 1986; 1987),
spiral structure
(e.g., Barbanis \& Woltjer 1967; Carlberg \& Sellwood 1985; Jenkins \& Binney 1990),
and stars,
tends to broaden
the velocity and spatial distributions of Galactic stars
(e.g., Wielen 1977; Nordstrom et al. 2004).

In order to roughly estimate the effect of dynamical heating
on NSs, Madau \& Blaes (1994)
applied the force-free diffusion equation
to the velocity and vertical distance distributions
of NSs.
They adopted the total diffusion coefficient, $C$
($=C_{U}+C_{V}+C_{W}$), measured by
Wielen (1977; $C=600$\,km$^{2}$\,s$^{-2}$\,Gyr$^{-1}$).
In this, $U$ corresponds to the radial direction,
and is positive in the direction
of the Galactic center;
$V$ points in the direction of circular rotation;
and $W$ is directed towards the North Galactic pole.
Applying heating, they found that the local density of NSs is smaller
by about 30\% relative to the case of no heating.
In the following we discuss this approach in light
of new observational data available.

\begin{figure}
\centerline{\includegraphics[width=8.5cm]{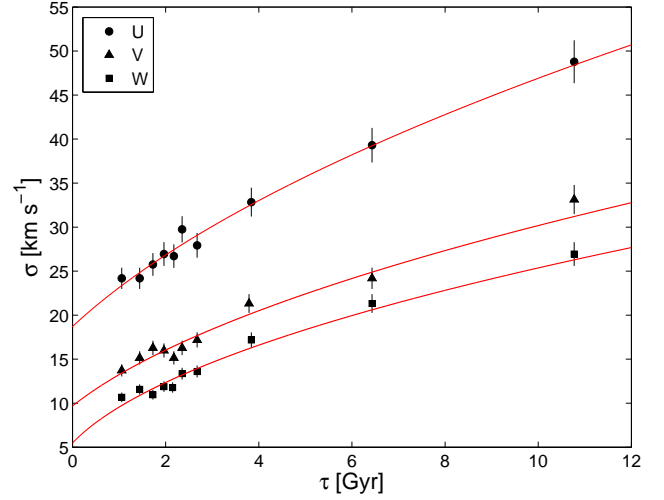}}
\caption{Stellar velocity dispersion ($\sigma$) along the $U$ (radial; {\it circles}),
$V$ (tangential; {\it triangles}) and $W$ (vertical; {\it squares})
Galactic directions
as a function of age since birth, $\tau$,
as measured by Nordstrom et al. (2004).
The lines represent the best fit of Equation~\ref{vel_diffus}
to the data. The best fit parameters are listed in Table~\ref{Tab:DiffusPar}.
\label{Fig:SigmaT}}
\end{figure}

We estimate the diffusion coefficient, $C$, using modern
data.
Nordstrom et al. (2004),
estimated the dynamical heating
by the Galactic disk and showed that it does not saturate
after some time as suggested by measurements based on
smaller samples (e.g., Quillen \& Garnett 2001; see also Aumer \& Binney 2009).
Approximating dynamical heating by a random walk process,
I fit the Nordstrom et al. (2004) measurements with the function
(e.g., Wielen 1977)
\begin{equation}
\sigma_{g} = (\sigma_{g,0}^{2} + C_{g}\tau)^{1/2},
\label{vel_diffus}
\end{equation}
where $\sigma_{g}$ is the velocity dispersion component
at time, $\tau$, since birth, $\sigma_{g,0}$ is the initial
velocity dispersion component and $g\in(U,V,W)$.
The data and the best fit curves are shown in Figure~\ref{Fig:SigmaT},
and the best fit parameters are listed in Table~\ref{Tab:DiffusPar}.
\begin{deluxetable}{lccc}
\tablecolumns{4}
\tablewidth{0pt}
\tablecaption{The Galactic disk diffusion coefficients}
\tablehead{
\colhead{Velocity component} &
\colhead{$\sigma_{g,0}$} &
\colhead{$C$} &
\colhead{$\chi^{2}/dof$} \\
\colhead{} &
\colhead{km~s$^{-1}$} &
\colhead{km$^{2}$~s$^{-2}$~Gyr$^{-1}$} &
\colhead{}
}
\startdata
 U    & $18.7\pm1.2$ & $185\pm19  $ & $3.1/8$  \\
 V    & $9.7\pm0.9 $ & $81.7\pm7.5$ & $9.3/8$  \\
 W    & $5.5\pm1.0 $ & $61.3\pm5.1$ & $11.0/8$
\enddata
\tablecomments{The diffusion coefficients of Galactic stars as obtained
by fitting Eq.~\ref{vel_diffus} with the measurements of Nordstrom et al. (2004).}
\label{Tab:DiffusPar}
\end{deluxetable}
The total diffusion coefficient
that I find, $C_{U}+C_{V}+C_{W}=328\pm21$\,km$^{2}$\,s$^{-2}$\,Gyr$^{-1}$,
is about half of the value found by Wielen (1977),
and used by Madau \& Blaes (1994).

Next,
the initial velocities estimated by
Narayan \& Ostriker (1990),
that was used by Madau \& Blaes (1994),
are considerably lower than the more
recent estimates (e.g., Arzoumanian et al. 2002).
As suggested by Eq.~\ref{vel_diffus},
diffusion affects mostly low velocity objects.
Therefore, dynamical heating is less important than estimated
by Madau \& Blaes (1994).

Furthermore, the approach taken by Madau \& Blaes (1994)
assumes that NSs are being affected by
diffusion at all times.
However, the scatterers are 
restricted to the Galactic plane.
NSs are born with high velocities and
spend $80\%$ to $90\%$ of the time at distances larger than
100\,pc from the Galactic plane (based on the results in \S\ref{sec:NSres}).
Therefore, they are less susceptible
to dynamical heating than disk stars.
Based on these three arguments, I conclude
that the importance of dynamical heating
was over estimated by Madau \& Blaes (1994) by at least an
order of a magnitude.

Nevertheless, dynamical heating may effect some of the slow moving objects,
but they are a minority among the Galactic NSs.
Eq.~\ref{vel_diffus} roughly suggests that
dynamical heating is important for NSs
with speeds smaller than about $(C\tau)^{1/2}$.
Even for $\tau=10$\,Gyr this gives 60\,km\,s$^{-1}$.
However, only $2.8\%$ and $4.5\%$ of the NSs
in the Arzoumanian et al. (2002)
and Faucher-Gigu\`{e}re \& Kaspi (2006),
initial velocity distributions, respectively,
have speeds smaller than 60\,km\,s$^{-1}$ at birth
(relative to their local standard of rest).

\section{Monte-Carlo simulations}
\label{sec:MC}

To solve for NS orbits I integrate
the equations of motion
\begin{equation}
\frac{d^{2}\vec{x}}{dt^{2}} = -\vec{\nabla} \Phi,
\label{eq:EqMotion}
\end{equation}
using a Livermore ordinary differential equation
solver\footnote{http://www.netlib.org/odepack/} (Hindmarsh 1983).
The integration is performed in a non-rotating, Cartesian coordinates system,
the origin of which is the Galactic center, and the $z$ axis is perpendicular
to the Galactic plane.

For each initial velocity distribution
(i.e., bimodal or unimodal; see \S\ref{sec:NSbirthdist}),
I simulated $3\times10^{6}$ NS orbits.
In each simulation I randomly drew the NSs birth times,
positions, and velocities, from the probability distributions described in
\S\ref{sec:NSbirthrate}, and \S\ref{sec:NSbirthdist}.
As explained before, $40\%$ of these NSs are disk-born,
and the rest are bulge-born.

At the end of each simulation I checked if the
integration conserved the total energy.
In cases in which the energy was not conserved to within $0.1\%$,
I reran the integration using the same initial conditions,
with a refined integration tolerance.
In the second (and final) iteration, the energy was conserved
to better than $2.8\%$ in all cases.

\section{Catalogue}
\label{sec:Cat}

The catalogue of initial
(i.e., at birth) and final
(i.e., current epoch)
simulated NSs space and velocity components
are listed in Table~\ref{tab:MC_v2g} for the bimodal initial velocity
distribution of Arzoumanian et al. (2002),
and in Table~\ref{tab:MC_v1e} for the unimodal initial velocity of
Faucher-Gigu\`{e}re \& Kaspi (2006).
The first column in each table indicates whether the
simulated NS belongs to the bulge population (code 0)
or disk population (code 1).
\begin{deluxetable*}{crrrrrrrr|rrrrrr}
\tablecolumns{15}
\tablewidth{0pt}
\tablecaption{NS Monte-Carlo orbital simulations using the bimodal initial velocity distribution}
\tablehead{
\multicolumn{9}{c}{Initial} &
\multicolumn{6}{c}{Final}   \\
\colhead{P} &
\colhead{Age} &
\colhead{$X$} &
\colhead{$Y$} &
\colhead{$Z$} &
\colhead{$\dot{X}$} &
\colhead{$\dot{Y}$} &
\colhead{$\dot{Z}$} &
\colhead{$V_{circ}$} &
\colhead{$X$} &
\colhead{$Y$} &
\colhead{$Z$} &
\colhead{$\dot{X}$} &
\colhead{$\dot{Y}$} &
\colhead{$\dot{Z}$} \\
\colhead{} &
\colhead{Gyr} &
\colhead{kpc} &
\colhead{kpc} &
\colhead{kpc} &
\colhead{kpc\,Gyr$^{-1}$} &
\colhead{kpc\,Gyr$^{-1}$} &
\colhead{kpc\,Gyr$^{-1}$} &
\colhead{kpc\,Gyr$^{-1}$} &
\colhead{kpc} &
\colhead{kpc} &
\colhead{kpc} &
\colhead{kpc\,Gyr$^{-1}$} &
\colhead{kpc\,Gyr$^{-1}$} &
\colhead{kpc\,Gyr$^{-1}$}
}
\startdata
0& 12.0&$-2.04$& $ 1.16$ & $-0.07$ & $ -77.7$ & $-194.6$ &$   8.8$&$208.7$&$ -0.71$&$   2.41$&$  -0.07$&$-181.0$&$  -70.7$&$   9.1$ \\
0& 12.0&$-0.58$& $ 1.57$ & $-0.27$ & $-199.1$ & $   0.1$ &$-376.7$&$206.7$&$ -1.40$&$  -8.23$&$  -5.08$&$  42.4$&$   25.4$&$ -70.3$ \\
0& 12.0&$ 0.60$& $ 1.57$ & $ 0.02$ & $ 531.9$ & $ 507.1$ &$ -60.9$&$206.6$&$477.31$&$ 404.37$&$  -5.35$&$-154.9$&$ -132.3$&$   2.3$ \\
0& 12.0&$-1.27$& $ 2.33$ & $ 0.05$ & $-140.6$ & $ -96.1$ &$ -53.5$&$211.6$&$ -0.20$&$   2.64$&$   0.06$&$-167.7$&$  -32.5$&$  52.4$ \\
0& 12.0&$-1.54$& $-2.95$ & $-0.18$ & $-137.7$ & $-209.2$ &$-425.3$&$218.2$&$ -1.31$&$  -1.29$&$ -19.70$&$ -10.4$&$   53.7$&$ 251.2$ 
\enddata
\tablecomments{Catalog of initial and final space positions and velocity components for $3\times10^{6}$ simulated NSs.
The NSs were simulated using the bimodal initial velocity distribution of Arzoumanian et al. (2002).
The velocity components are given in kpc\,Gyr$^{-1}$,
relative to the non-rotating Galaxy.
In order to convert kpc\,Gyr$^{-1}$ to km\,s$^{-1}$,
divide it by $1.0227$.
P is the population type: 0 for bulge-born NS; 1 for disk-born NS.
$V_{circ}$ refers to the Galaxy rotation speed at the projected location,
on the Galactic disk, in which
the NS was born. The initial (and final) velocity components
are given relative to a non-rotating galaxy (inertial reference frame).
The numbers in this table are rounded in order to fit into the page.
This table is published in its entirety in the electronic edition of this
paper.
A portion of the full table is shown here for guidance regarding its form and content.}
\label{tab:MC_v2g}
\end{deluxetable*}
\begin{deluxetable*}{crrrrrrrr|rrrrrr}
\tablecolumns{15}
\tablewidth{0pt}
\tablecaption{NS Monte-Carlo orbital simulations using the unimodal initial velocity distribution}
\tablehead{
\multicolumn{9}{c}{Initial} &
\multicolumn{6}{c}{Final}   \\
\colhead{P} &
\colhead{Age} &
\colhead{$X$} &
\colhead{$Y$} &
\colhead{$Z$} &
\colhead{$\dot{X}$} &
\colhead{$\dot{Y}$} &
\colhead{$\dot{Z}$} &
\colhead{$V_{circ}$} &
\colhead{$X$} &
\colhead{$Y$} &
\colhead{$Z$} &
\colhead{$\dot{X}$} &
\colhead{$\dot{Y}$} &
\colhead{$\dot{Z}$} \\
\colhead{} &
\colhead{Gyr} &
\colhead{kpc} &
\colhead{kpc} &
\colhead{kpc} &
\colhead{kpc\,Gyr$^{-1}$} &
\colhead{kpc\,Gyr$^{-1}$} &
\colhead{kpc\,Gyr$^{-1}$} &
\colhead{kpc\,Gyr$^{-1}$} &
\colhead{kpc} &
\colhead{kpc} &
\colhead{kpc} &
\colhead{kpc\,Gyr$^{-1}$} &
\colhead{kpc\,Gyr$^{-1}$} &
\colhead{kpc\,Gyr$^{-1}$}
}
\startdata
1&  2.3&$-7.05$&$-6.69$&$ 0.07$&$-258.0$&$-232.1$&$ -55.2$&$220.2$&$ -39.41$&$  -3.21$&$  -5.51$&$   24.9$&$    4.3$&$    8.8$ \\
0& 12.0&$ 0.38$&$ 0.88$&$-0.01$&$-124.3$&$-176.5$&$ 134.8$&$226.7$&$   0.35$&$   0.10$&$   0.60$&$  213.6$&$  178.4$&$   79.1$ \\
1&  2.9&$-0.69$&$ 5.40$&$ 0.07$&$-124.5$&$ 122.9$&$ 294.4$&$227.1$&$  -6.97$&$  -5.47$&$ -11.00$&$   34.2$&$  -57.3$&$  105.9$ \\
1&  0.6&$-2.02$&$-1.49$&$-0.02$&$-510.5$&$-374.6$&$   5.6$&$210.2$&$-140.81$&$-103.14$&$  -0.48$&$ -139.0$&$ -101.7$&$   -0.8$ \\
0& 12.0&$-2.34$&$-2.96$&$-0.02$&$ 178.0$&$-374.0$&$-108.7$&$221.6$&$  -7.62$&$ -17.92$&$   2.54$&$  115.5$&$   87.9$&$  -21.8$
\enddata
\tablecomments{Like Table~\ref{tab:MC_v2g}, but for the
unimodal initial velocity distribution of Faucher-Gigu\`{e}re \& Kaspi (2006).}
\label{tab:MC_v1e}
\end{deluxetable*}

In addition, 
catalogue of simulated NS positions, distances,
radial velocities, and proper motions
for an observer located at the ``solar circle'', moving around
the Galactic center with a velocity of 220\,km\,s$^{-1}$
is given in tables~\ref{tab:SunCentric2} and \ref{tab:SunCentric1}.
The solar circle is defined to be
on the Galactic plane (i.e., $z=0$\,kpc)
at the distance of the Sun from the Galactic center
($R_{\odot}=8.0$\,kpc; Ghez et al. 2008).
The catalogue
in Tables~\ref{tab:SunCentric2} and \ref{tab:SunCentric1}
are based on the initial velocity distribution
of Arzoumanian et al. (2002) and
the Faucher-Gigu\`{e}re \& Kaspi (2006), respectively.
I note that the radial velocity and proper motions
are calculated for a static observer with respect to the
Local Standard of Rest (LSR;
i.e., the solar motion
with respect to the LSR is neglected).

In order to reduce Poisson errors in
the local properties of NSs (i.e., density; sky surface density),
tables~\ref{tab:SunCentric2} and \ref{tab:SunCentric1}
were produced by calculating the positions
of the $3\times10^{6}$ simulated NSs
in tables~\ref{tab:MC_v2g} and \ref{tab:MC_v1e}, respectively,
from 100 random locations on the solar circle.
Hence, tables~\ref{tab:SunCentric2} and \ref{tab:SunCentric1}
list $3\times10^{8}$ simulated NSs.
\begin{deluxetable*}{crrrrrrr}
\tablecolumns{8}
\tablewidth{0pt}
\tablecaption{Catalog of simulated NSs as observed from the LSR - based on the bimodal initial velocity distribution}
\tablehead{
\colhead{P} &
\colhead{Age} &
\colhead{$l$} &
\colhead{$b$} &
\colhead{dist} &
\colhead{$\mu_{l}$} &
\colhead{$\mu_{b}$} &
\colhead{RV} \\
\colhead{} &
\colhead{Gyr} &
\colhead{deg} &
\colhead{deg} &
\colhead{kpc} &
\colhead{$''$\,yr$^{-1}$} &
\colhead{$''$\,yr$^{-1}$} &
\colhead{km\,s$^{-1}$}
}
\startdata
0 & 12.0000 &$  354.72497$&$   -0.39865$&$       10.0003$&$    -0.0064166$&$    0.0001654$&$  -149.7$ \\
0 & 12.0000 &$   72.14501$&$  -38.10227$&$        8.2258$&$    -0.0025465$&$   -0.0041258$&$   -93.8$ \\
0 & 12.0000 &$  273.07904$&$   -0.49003$&$      625.9548$&$    -0.0000573$&$    0.0000017$&$   340.7$ \\
0 & 12.0000 &$  351.53160$&$    0.34229$&$        9.9375$&$    -0.0057996$&$    0.0011040$&$  -124.8$ \\
0 & 12.0000 &$   13.81616$&$  -68.84354$&$       21.1283$&$    -0.0015999$&$    0.0004210$&$  -247.1$
\enddata
\tablecomments{Catalog of Galactic longitudes ($l$), latitudes ($b$),
distances, proper motions in Galactic longitude ($\mu_{l}$) and latitude ($\mu_{b}$),
and radial velocities (RV) from a point on the solar circle
(e.g., the LSR) of $3\times10^{8}$ simulated NSs.
The velocities and proper motions do~not include the motion
of the Sun relative to the LSR.
The proper motions are given in the Galactic coordinate system.
The catalog was generated by calculating the positions
of the $3\times10^{6}$ NSs in Table~\ref{tab:MC_v2g}
(i.e., assuming the Arzoumanian et al. [2002]
initial velocity distribution)
as observed from 100 random points on the solar circle.
The conversion of space velocity to proper motion and radial velocity
was carried out using the inverse of Eq. 3.23-3 in Seidelmann (1992, p. 121).
This table is published in its entirety in the electronic edition of
this paper.
A portion of the full table is shown here for guidance
regarding its form and content.
}
\label{tab:SunCentric2}
\end{deluxetable*}
\begin{deluxetable*}{crrrrrrr}
\tablecolumns{8}
\tablewidth{0pt}
\tablecaption{Catalog of simulated NSs as observed from the LSR - based on the unimodal initial velocity distribution}
\tablehead{
\colhead{P} &
\colhead{Age} &
\colhead{$l$} &
\colhead{$b$} &
\colhead{dist} &
\colhead{$\mu_{l}$} &
\colhead{$\mu_{b}$} &
\colhead{RV} \\
\colhead{} &
\colhead{Gyr} &
\colhead{deg} &
\colhead{deg} &
\colhead{kpc} &
\colhead{$''$\,yr$^{-1}$} &
\colhead{$''$\,yr$^{-1}$} &
\colhead{km\,s$^{-1}$}
}
\startdata
1 &  2.2753 &$  237.98450$&$   -8.95638$&$       35.3986$&$     0.0008045$&$    0.0002084$&$   166.7$ \\
0 & 12.0000 &$    2.28569$&$    4.42334$&$        7.8361$&$    -0.0014499$&$    0.0016472$&$   212.2$ \\
1 &  2.8584 &$  306.10398$&$  -45.37962$&$       15.4574$&$    -0.0018509$&$    0.0033513$&$    96.8$ \\
1 &  0.6194 &$  262.62769$&$   -0.15823$&$      173.4066$&$    -0.0001141$&$    0.0000002$&$   334.3$ \\
0 & 12.0000 &$  308.68736$&$    6.21614$&$       23.4368$&$     0.0000392$&$   -0.0003617$&$   171.9$
\enddata
\tablecomments{Like Table~\ref{tab:SunCentric1}, but for the
unimodal initial velocity distribution of Faucher-Gigu\`{e}re \& Kaspi (2006).
}
\label{tab:SunCentric1}
\end{deluxetable*}
%
%

\section{Statistical properties}
\label{sec:NSres}

In this section I present 
the space and velocity distributions
of simulated NSs, at the current epoch.
In \S\ref{sec:GenProp} I discuss the
overall distribution of NSs in the Galaxy,
while in \S\ref{sec:ObsProp} I discuss their 
statistical properties as observed
from the LSR.
Additional specific statistical properties of
these objects are discussed in
Ofek et al. (2009)
in the context of the long-duration radio transients
(Bower et al. 2007; Kida et al. 2008).
The results presented in this section
assume there are $N_{9}=N/10^{9}$ NSs in the Galaxy,
where $N$ is the number of NSs in the Galaxy,
of which $40\%$ were born in the disk and $60\%$ in the bulge.

\subsection{Overall properties}
\label{sec:GenProp}

NSs are born with large space velocities,
which are typically of the order of the escape
velocity from the Galaxy.
These are presumably the result
of kick velocities due
to asymmetric supernovae explosions (e.g., Blaauw 1961 ; Lai et al. 2006).
Therefore, it is expected that a large fraction
of the Galactic NSs will be unbounded to the
Milky Way gravitational potential
and some may be found at very large distances.

\begin{figure}
\centerline{\includegraphics[width=8.5cm]{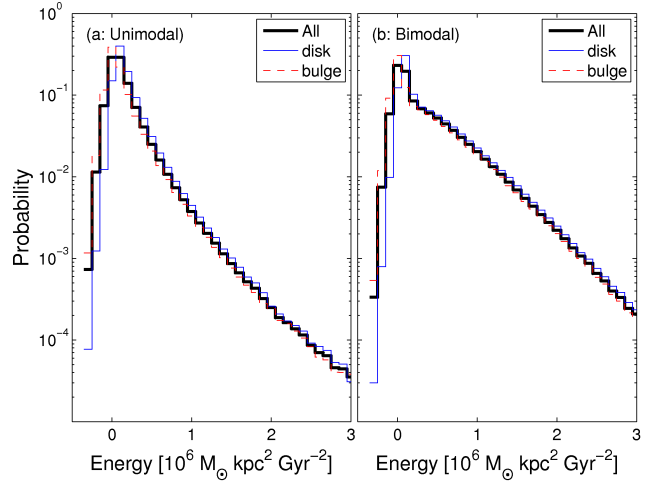}}
\caption{The total energy distribution
of Galactic-born NSs.
The energy does~not include external potentials (i.e., other galaxies),
and assumes all NSs have a mass of 1.4\,M$_{\odot}$.
The different lines shows
the distribution of all the NSs ({\it solid thick black} lines);
only disk-born NSs ({\it solid thin} lines);
and only bulge-born NSs ({\it dashed} lines).
\label{fig:NS_EnergyDist}}
\end{figure}

In Figure~\ref{fig:NS_EnergyDist}, I show the distribution of the
total (kinetic $+$ potential) energy of the simulated NSs:
$M_{NS} ( v^{2}/2 + \Phi )$,
where the NSs mass was set to $M_{NS}=1.4$\,M$_{\odot}$.
Panel (a) shows the energy distribution based on the 
simulations using the unimodal
initial velocity distribution of Faucher-Gigu\`{e}re \& Kaspi (2006),
while panel (b) is for the bimodal distribution
of Arzoumanian et al. (2002).
In each panel, the {\it thick black solid} line represents the
entire NS population,
the {\it thin solid} line shows the disk-born NSs, and the {\it dashed} line 
is for bulge-born NSs.
Using the approximation that all the NSs with negative energies
are gravitationally bounded to the Galaxy (neglecting heating),
in Table~\ref{Tab:unboundedNS} I give
the fractions of NSs bounded to the Galactic gravitational potential.
\begin{deluxetable}{lccc}
\tablecolumns{4}
\tablewidth{0pt}
\tablecaption{Fraction of NSs gravitationally bounded to the Galaxy}
\tablehead{
\colhead{Initial velocity\tablenotemark{a}} &
\colhead{All} &
\colhead{disk} &
\colhead{bulge}
}
\startdata
A2002   & 0.38 & 0.16 & 0.52 \\
FK2006  & 0.30 & 0.13 & 0.41
\enddata
\tablenotetext{a}{Initial velocity distribution used in the simulations,
where A2002 corresponds to
Arzoumanian et al. (2002), and FK2006 to Faucher-Gigu\`{e}re \& Kaspi (2006).}
\label{Tab:unboundedNS}
\end{deluxetable}

Given the large fraction of gravitationally-unbounded NSs,
it is expected that some NSs may be found at very large distances, $r$,
from the Galactic center.
I find that
about $12\%$ and $35\%$ of the NSs born in the Galaxy
are currently at distances larger than 1\,Mpc
from the Galactic center,
for the unimodal and bimodal
initial velocity distributions, respectively.
For $1$\,Mpc$<r<10$\,Mpc, I find that the density of NSs
as a function of $r$ is about:
$1.9\times10^{-5}(r/1 {\rm kpc})^{-2.4}N_{9}$\,pc$^{-3}$
and
$3.6\times10^{-6}(r/1 {\rm kpc})^{-2.1}N_{9}$\,pc$^{-3}$,
for the bimodal and unimodal initial velocity distributions, respectively.
Finally, I find that some Milky Way born NSs may be
at distances as large as 30 to 40\,Mpc from the Galaxy.
I note that the local density, in our Galaxy, of NSs born in other galaxies,
is of the order of $10^{-11}N_{9}$\,pc$^{-3}$.
This was estimated by calculating the density
in the the Milky Way, of NSs born in each galaxy found within 10\,Mpc.
For this, I used a version of the Tully (1988) nearby galaxy catalog
(Ofek 2007)
where the total number of NSs in each galaxy was normalized by its
total $B$-band magnitude, relative to Milky Way.

\begin{figure}
\centerline{\includegraphics[width=8.5cm]{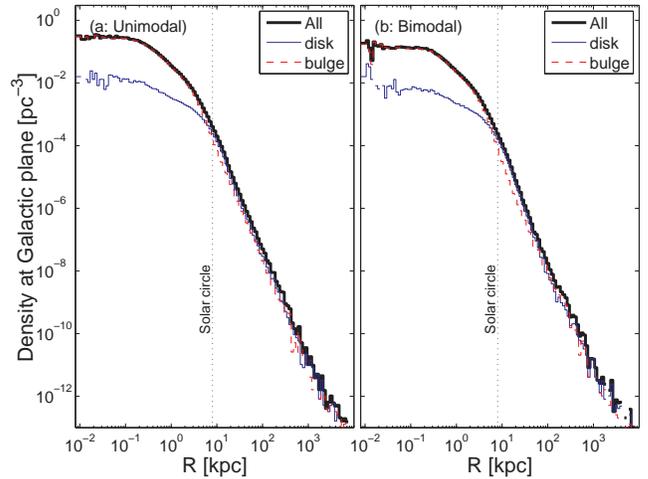}}
\caption{Density of NSs at the Galactic plane,
as measured at the current epoch.
See Fig.~\ref{fig:NS_EnergyDist}
for details regarding line types.
The density is calculated by counting the NSs within
100\,pc from the Galactic plane, dividing it
by the appropriate volume in each bin,
and assuming that the total number of NSs in the Galaxy is $10^{9}$.
For the initial velocity distribution of
Faucher-Gigu\`{e}re \& Kaspi (2006; panel a),
the NSs density at the
solar circle
(distance of 8.0\,kpc from the Galactic center; Ghez et al. 2008)
is
$4.0\times10^{-4}N_{9}$\,pc$^{-3}$,
of which $60\%$ are disk born and $40\%$ are bulge born.
The NSs density at the Galactic center is:
$3\times10^{-1}N_{9}$\,pc$^{-3}$,
of which $5\%$ are disk born and $95\%$ are bulge born.
For the initial velocity distribution of
Arzoumanian et al. (2002; panel b),
the density at the
solar circle is
$2.4\times10^{-4}N_{9}$\,pc$^{-3}$,
of which $66\%$ are disk born and $34\%$ are bulge born.
The density at the Galactic center is:
$2\times10^{-1}N_{9}$\,pc$^{-3}$,
of which $7\%$ are disk born and $93\%$ are bulge born.
I note that the fluctuations seen in small and large radii
are due to Poisson noise.
\label{fig:NS_RadialDensity}}
\end{figure}

\begin{figure}
\centerline{\includegraphics[width=8.5cm]{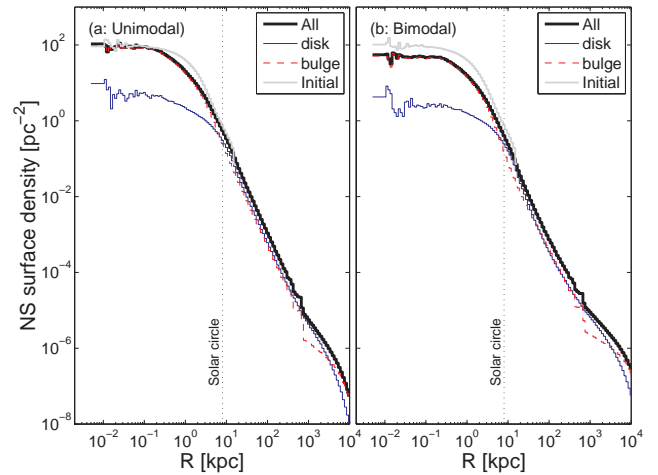}}
\caption{The surface density of NSs projected on the Galactic plane,
as measured at the current epoch.
See Fig.~\ref{fig:NS_EnergyDist} for details.
In addition the {\it gray} lines show the initial surface density distributions.
For the initial velocity distribution of
Faucher-Gigu\`{e}re \& Kaspi (2006; panel a),
the NSs surface density at the
solar circle is
$0.6N_{9}$\,pc$^{-2}$
of which $54\%$ are disk born NSs, and $46\%$ are bulge-born NSs.
The surface density at the Galactic center is
$\sim90N_{9}$\,pc$^{-2}$
of which about $9\%$ are disk born NSs, and the rest are bulge-born NSs.
For the initial velocity distribution of
Arzoumanian et al. (2002; panel b),
the NSs density at the
solar circle is
$0.4N_{9}$\,pc$^{-2}$
of which $61\%$ are disk born NSs, and $39\%$ are bulge-born NSs.
The corresponding surface density at the Galactic center is
$\sim50N_{9}$\,pc$^{-2}$
of which $\sim6\%$ are disk born NSs, and the rest are bulge-born NSs.
As before, the plotted densities assume the total number of NSs
is $10^{9}$.
\label{fig:NS_RadialSurfaceDensity}}
\end{figure}

In Figures~\ref{fig:NS_RadialDensity} and \ref{fig:NS_RadialSurfaceDensity}
I show the density of NSs on the Galactic plane, and the surface density
of NSs projected on the Galactic plane, respectively,
as a function of distance from the Galactic center.
The notations are the same as in
Figure~\ref{fig:NS_EnergyDist}.
In addition, in Fig.~\ref{fig:NS_RadialSurfaceDensity},
the {\it gray solid} line represents the initial
surface distribution of all NSs.
Furthermore, the vertical {\it dashed} lines mark the distance
of the Sun from the Galactic center,
$R_{\odot}=8.0$\,kpc (Ghez et al. 2008).
As noted before, I refer to this distance from
the Galactic center, when located on the Galactic plane,
as the solar circle.
The densities in the solar circle and Galactic center
are listed in the figure captions.
I note that the fluctuations seen in small and large radii
are due to Poisson noise.

\begin{figure}
\centerline{\includegraphics[width=8.5cm]{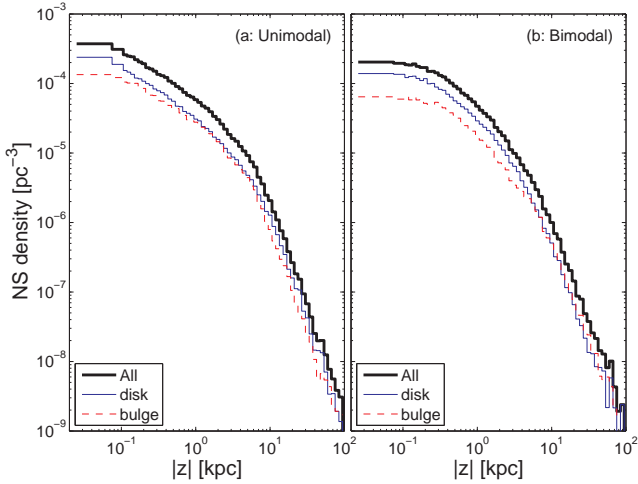}}
\caption{The density of NSs as a function of height, $z$,
above/below the Galactic plane
as measured on the solar circle 
at the current epoch.
Line types are like those in Fig.~\ref{fig:NS_EnergyDist}.
For the initial velocity distribution of
Faucher-Gigu\`{e}re \& Kaspi (2006; panel a),
$50\%$ ($90\%$) of the NSs are found within 0.9\,kpc (5.6\,kpc)
from the Galactic plane.
For the initial velocity distribution of
Arzoumanian et al. (2002; panel b),
$50\%$ ($90\%$) of the NSs are found within 0.7\,kpc (4.4\,kpc)
from the Galactic plane.
\label{fig:NS_Gal_zDistSolarCirc}}
\end{figure}

In Figure~\ref{fig:NS_Gal_zDistSolarCirc} I show
the NSs density
as a function of height above or below the Galactic plane,
as measured at the solar circle.
This was calculated by counting the number
of simulated NSs with
distance from the Galactic center,
projected on the Galactic plane, of
$8.0\pm0.5$\,kpc.
The line scheme is the same as in
Figures~\ref{fig:NS_EnergyDist}.
At the solar circle the
scale height\footnote{Scale height is defined as the height at which the density drops by $1/e$.}
of NSs is about $0.6$\,kpc
and $0.3$\,kpc for the
Arzoumanian et al. (2002)
and
Faucher-Gigu\`{e}re \& Kaspi (2006) 
initial velocity distributions, respectively.

\begin{figure}
\centerline{\includegraphics[width=8.5cm]{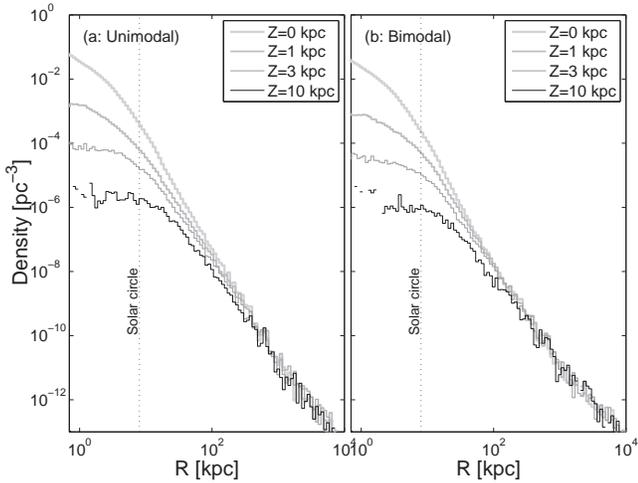}}
\caption{The density of NSs in Galactic heights of
0\,kpc, 1\,kpc, 3\,kpc, and 10\,kpc
above/below the Galactic plane,
as a function of the projected (on the Galactic plane)
distance from the Galactic
center, $R$, and as measured at the current epoch.
Panel (a) shows the densities calculated using
the initial velocity distribution of
Faucher-Gigu\`{e}re \& Kaspi (2006).
For this model, at the solar circle,
the densities are about
$4\times10^{-4}N_{9}$\,pc$^{-3}$,
$6\times10^{-5}N_{9}$\,pc$^{-3}$,
$2\times10^{-5}N_{9}$\,pc$^{-3}$, and
$2\times10^{-6}N_{9}$\,pc$^{-3}$,
at Galactic heights of
0\,kpc, 1\,kpc, 3\,kpc, and 10\,kpc, respectively.
Panel (b) shows the densities calculated using
the initial velocity distribution of
Arzoumanian et al. (2002).
For this model, at the solar circle,
the densities are about
$2\times10^{-4}N_{9}$\,pc$^{-3}$,
$5\times10^{-5}N_{9}$\,pc$^{-3}$,
$1\times10^{-5}N_{9}$\,pc$^{-3}$, and
$1\times10^{-6}N_{9}$\,pc$^{-3}$,
at Galactic heights of
0\,kpc, 1\,kpc, 3\,kpc, and 10\,kpc, respectively.
\label{fig:NS_RadialDensityZ}}
\end{figure}

Shown in Fig.~\ref{fig:NS_RadialDensityZ} are the
space densities of NSs as a function of projected (on the Galactic plane)
distance from the Galactic center,
for several different Galactic height:
0\,kpc; 1\,kpc; 3\,kpc; and 10\,kpc.
The densities are calculated in slices,
parallel to the Galactic plane,
with semi-width of 0.1\,kpc.

\begin{figure}
\centerline{\includegraphics[width=8.5cm]{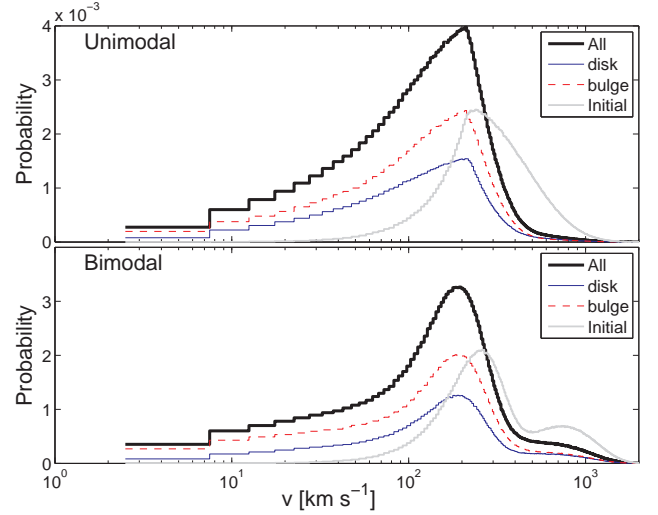}}
\caption{The initial and final speed probability distributions
of the simulated NSs, relative to an inertial reference frame.
Line types are like those in Fig.~\ref{fig:NS_RadialSurfaceDensity}.
The probabilities are calculated for 1\,km\,s$^{-1}$ bins.
Panel (a) shows the results for the unimodal initial velocity
distribution, while panel (b)
for the bimodal initial velocity distribution.
\label{fig:NS_SpeedDist}}
\end{figure}

Finally, in Figure~\ref{fig:NS_SpeedDist}
I show the initial and final speed distributions
as measured relative to an inertial reference frame
(contrary to Fig.~\ref{fig:InitVelDist_Compare} which shows
the initial speed distribution relative to
a rotating reference frame).
The probabilities in this Figure are shown per 1\,km\,s$^{-1}$ bins.
The line scheme is again like the one used in
Fig.~\ref{fig:NS_RadialSurfaceDensity}.
As expected, the typical speeds of NSs
decrease with time as they, on average,
increase their distances from the Galaxy
and lose kinetic energy.
For the unimodal initial velocity distribution,
at the current epoch I estimate that
about $17\%$, $53\%$, and $99\%$ of the NSs have speeds below
100, 200, and 1000\,km\,s$^{-1}$, respectively.
For the bimodal initial velocity distribution,
at the current epoch I estimate that
about $11\%$, $39\%$, and $94\%$ of the NSs have speeds below
100, 200, and 1000\,km\,s$^{-1}$, respectively.

\subsection{Properties observed from the LSR}
\label{sec:ObsProp}

In this subsection I discuss:
(i) the expectancy distance of the nearest NS to the Sun;
(ii) the number of young NSs in the solar neighborhood;
(iii) the proper motion distribution of nearby NSs;
and (iv) the all-sky distribution of Galactic NSs.

The probability to find a NS within distance $d$
from the Sun is given by
\begin{equation}
P_{<d} = 1-\exp(-\frac{4}{3}\pi \rho d^{3}).
\label{Pd}
\end{equation}
Given the local density of NSs, $\rho$,
that I have found in \S\ref{sec:GenProp},
this implies that
the expectancy distance of the nearest NS
is about $8.8N_{9}^{-1/3}$\,pc and $7.5N_{9}^{-1/3}$\,pc
for the
Arzoumanian et al. (2002)
and
Faucher-Gigu\`{e}re \& Kaspi (2006) distributions, respectively.

For the bimodal initial velocity distribution
(Table~\ref{tab:SunCentric2}),
within 1\,kpc from the Sun, 
63\% of the NSs are disk-born,
and there are about $220 N_{9}$ ($900 N_{9}$) NSs
younger than 1\,Myr (10\,Myr).
On the other hand,
for the unimodal initial velocity distribution (Table~\ref{tab:SunCentric1}),
within 1\,kpc from the Sun,
57\% of the NSs are disk-born, and 
there are about $190 N_{9}$ ($930 N_{9}$) NSs
younger than 1\,Myr (10\,Myr).

\begin{figure}
\centerline{\includegraphics[width=8.5cm]{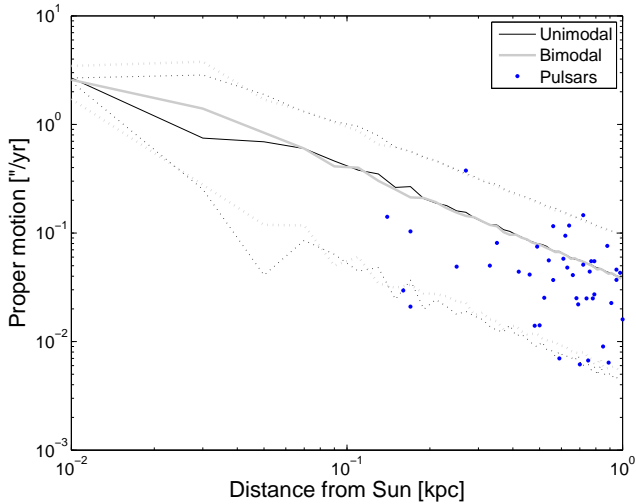}}
\caption{The median total proper motion
({\it solid} lines)
of simulated NSs as a function of their distance from
an observer located on the solar circle.
The {\it dotted} lines show the lower and upper $95$-percentiles
of the proper motion distributions.
The {\it black} lines represent the unimodal initial velocity distribution
and the {\it gray} lines are for the bimodal initial velocity distribution.
In addition, the {\it dots} show the observed proper motion of known pulsars.
\label{fig:NS_PM_Dist}}
\end{figure}

In Figure~\ref{fig:NS_PM_Dist}, I show the median total proper motion
({\it solid} lines)
of simulated NSs as a function of their distance from
an observer located on the solar circle.
The {\it dotted} lines show the lower and upper $95$-percentiles
of the proper motion distributions.
The {\it black} lines represent the unimodal initial velocity distribution
and the {\it gray} lines are for the bimodal initial velocity distribution.
In addition, the {\it dots} show the observed proper motions of known
pulsars\footnote{Pulsar distances and proper motions obtained from: http://www.atnf.csiro.au/research/pulsar/psrcat/.}
(Manchester et al. 2005).
%
%
\begin{figure*}
\centerline{\includegraphics[width=16cm]{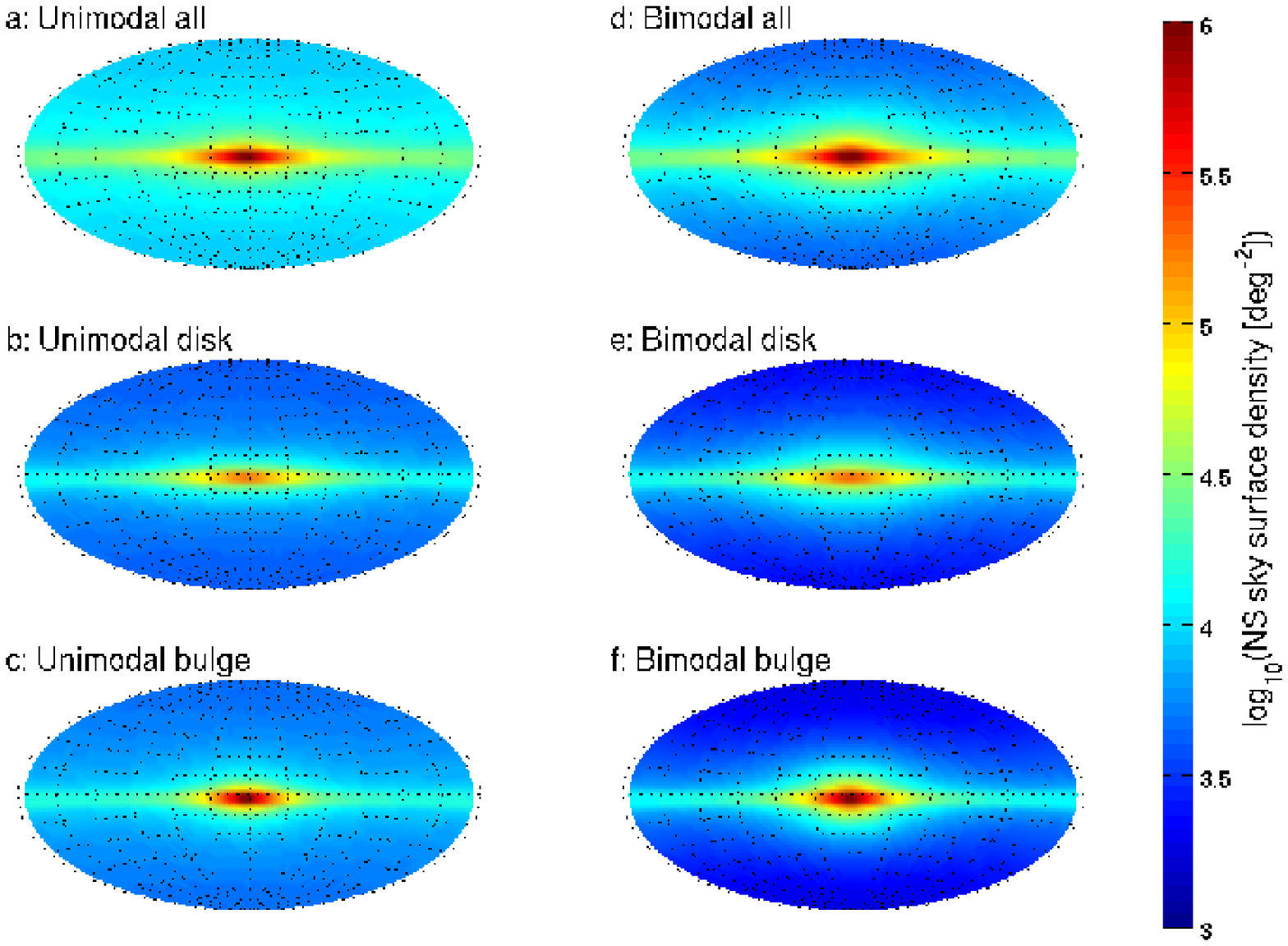}}
\caption{Sky surface density distribution of NSs, at the current epoch,
for an observer located on the solar circle.
The maps
are presented using the Aitoff equal area projection. The grids represent
Galactic coordinates with $15^{\circ}$ spacing.
The panels in the left column are for the
unimodal initial velocity distribution
of Faucher-Gigu\`{e}re \& Kaspi (2006),
while the right column is for the bimodal initial
velocity distribution of Arzoumanian et al. (2002).
The upper row is assuming $10^{9}$ NSs of which $60\%$
are bulge born and $40\%$ are disk born.
The middle row is for the disk born NSs,
assuming there are $4\times10^{8}$ of them in the Galaxy,
while the bottom row is for the bulge born NSs,
assuming there are $6\times10^{8}$ of them in the Galaxy.
\label{fig:NS_SkyDistribution}}
\end{figure*}
I note that the largest total proper motion in
Table~\ref{tab:SunCentric2} (i.e., bimodal initial velocity distribution)
is $4.6''$\,yr$^{-1}$, and that $\sim170 N_{9}$ NSs are expected to
have proper motion in excess of $1''$\,yr$^{-1}$.
In Table~\ref{tab:SunCentric1}, the largest proper motion
is $3.3''$\,yr$^{-1}$,
and about $240 N_{9}$ NSs are expected to
have proper motion in excess of $1''$\,yr$^{-1}$.

In Fig.~\ref{fig:NS_SkyDistribution}a--f
I show the sky surface density of NSs at the current epoch,
for an observer located at the solar circle.
Panels (a)--(c) are for the unimodal initial velocity distribution,
while panels (d)--(f) are for the bimodal initial velocity distribution.
Panels (a) and (d) show the distribution for
all the NSs (i.e., disk- and bulge-born populations),
panels (b) and (e) for the disk-born population,
and panels (c) and (f) for the bulge-born population.
The maps are shown in the Aitoff equal-area projection
and Galactic coordinate system, where
the Galactic center is at the center of each map.
The surface densities are normalized assuming that there are $4\times10^{8}$
disk-born NSs, and $6\times10^{8}$ bulge-born NSs.

At the positions with the lowest surface density of NSs,
the Poisson errors due to the limited statistics
are smaller than about $10\%$.
I find that the minimum surface density is attained
at the direction of the Galactic poles, and
is about $3900N_{9}$\,deg$^{-2}$ and $8100N_{9}$\,deg$^{-2}$
for the unimodal and bimodal initial velocity distributions,
respectively.
The maximum surface density is at the direction of the Galactic center
and it is about $1.4\times10^{6}N_{9}$\,deg$^{-2}$
and $1.1\times10^{6}N_{9}$\,deg$^{-2}$,
for the unimodal and bimodal initial velocity distributions,
respectively.
I note that the differences between the sky surface densities
resulting from the two initial velocity distributions,
are as large as about $65\%$.

\section{Summary}
\label{sec:Disc}

The Milky Way's NSs content, and in particular the NS
space and velocity distributions
are important for searching nearby isolated old NSs,
and studying any ongoing activity from such objects.
As a tool for such studies,
I present a mock catalogue of simulated isolated old
NSs spatial positions and velocities,
at the current epoch.

The catalogue
were constructed by integrating the equations of motion
of simulated NSs in the Galactic potential.
The simulations include two populations of NSs,
one in which the NSs were born
in the Galactic bulge about 12\,Gyr ago, and the second
population in which NSs are being born at the Galactic disk
at a constant rate, starting 12\,Gyr ago.
The combined NS population assumes that $60\%$
of the NSs originated in the bulge and the rest
in the disk.
Although we do~not know what is the exact number
of Galactic NSs, and what was their position-dependent
birth rate, these two populations provide
a wide range of initial conditions.

I generated two catalogue of simulated NSs.
Each catalog contains $3\times10^{6}$ objects.
The two catalogue utilize different initial velocity
distributions of NSs.
One catalog (Table~\ref{tab:MC_v2g})
uses the initial velocity distribution of
Arzoumanian et al. (2002),
while the other (Table~\ref{tab:MC_v1e})
uses the initial velocity distribution of
Faucher-Gigu\`{e}re \& Kaspi (2006).
Also derived are catalogue of simulated NS positions
and proper motions with respect to an observer
found at the solar circle
(tables~\ref{tab:SunCentric2} and \ref{tab:SunCentric1}).

The space distribution at the current epoch
obtained by the two initial velocity distributions
implemented here,
are somewhat different.
For example, I find that the
resulting sky surface density, based on the two
different initial velocity distributions,
differs by up to $65\%$.
The main differences between the two velocity distributions
are in the low- and high-tails of the NSs velocity distribution
(see Fig.~\ref{fig:InitVelDist_Compare}).

\begin{deluxetable}{lccl}
\tablecolumns{4}
\tablewidth{0pt}
\tablecaption{Comparison with previous works}
\tablehead{
\colhead{$\rho/N_{9}$} &
\colhead{$Z_{1/2}$} &
\colhead{mode$(p[v])$\tablenotemark{a}} &
\colhead{Reference} \\
\colhead{pc$^{-3}$} &
\colhead{kpc} &
\colhead{km\,s$^{-1}$} &
\colhead{}
}
\startdata
$2.4\times10^{-4}$  &0.42  & 240      & This Work (disk$+$bulge) A2002\tablenotemark{b} \\
$4.0\times10^{-4}$  &0.20  & 200      & This Work (disk$+$bulge) FK2006\tablenotemark{c} \\
$1.4\times10^{-3}$  &0.20  &          & Paczynski (1990) \\
$7.5\times10^{-4}$  &0.27  &  43      & Blaes \& Madau (1993) \\
$5.3\times10^{-4}$  &0.50  &  69      & Madau \& Blaes (1994) \\
$4\times10^{-4}$    &      & 140      & Perna et al. (2003) \\ 
$5\times10^{-4}$    &0.4   &          & Popov et al. (2005)
\enddata
\tablenotetext{a}{Mode is the most probable value of the distribution, and the speeds are measured relative to an inertial reference frame.}
\tablenotetext{b}{Using the bimodal initial velocity distribution of Arzoumanian et al. (2002).}
\tablenotetext{c}{Using the unimodal double-sided exponential initial velocity distribution of Faucher-Gigu\`{e}re \& Kaspi (2006).}
\tablecomments{The density, $\rho$,
at the solar neighborhood
is calculated assuming there are $10^{9}$ NSs in the Galaxy.
$Z_{1/2}$ is the height above the Galactic plane, in which
the NS density drops to $1/2$ its value on the Galactic plane,
calculated at the solar circle.
The density from Madau \& Blaes (1994) is calculated assuming that diffusion
operats for 5\,Gyr (see \S\ref{sec:DynHeat}).
The initial velocity distribution used by Paczynski (1990),
Blaes \& Madau (1993), and Madau \& Blaes (1994)
is from Narayan \& Ostriker (1990).
The initial velocity distribution in Perna et al. (2003)
is from Cordes \& Chernoff (1998).
}
\label{tab:Comp}
\end{deluxetable}
The space and velocity distributions of Galactic NSs
were estimated by several past works.
In Table~\ref{tab:Comp}, I compare some of the basic
results of the simulations presented here
(e.g., local density of NSs; scale height),
with the ones obtained by previous efforts.

\acknowledgments
I thank
Re'em Sari, Ehud Nakar, Orly Gnat, Avishay Gal-Yam, Sterl Phinney,
and Mansi Kasliwal for valuable discussions.
I also thank an anonymous referee for valuable comments.
Support for program number HST-GO-11104.01-A was provided by NASA through
a grant from the Space Telescope Science Institute, which is
operated by the Association of Universities for Research in
Astronomy, Incorporated, under NASA contract NAS5-26555.

\end{document}